Observational Evidence for Tidal Destruction of Exoplanets

by

Brian Jackson[1], Rory Barnes[1,2,3], & Richard Greenberg[1]

[1]Lunar and Planetary Laboratory
University of Arizona
1629 E University Blvd, Tucson AZ 85721-0092

[2]Astronomy Department
University of Washington
Box 351580, Seattle WA

[3]Virtual Planetary Laboratory

Running Title: Evidence for Tidal Destruction of Exoplanets

Abstract

The distribution of the orbits of close-in exoplanets shows evidence for on-going removal and destruction by tides. Tides raised on a planet's host star cause the planet's orbit to decay, even after the orbital eccentricity has dropped to zero. Comparison of the observed orbital distribution and predictions of tidal theory show good qualitative agreement, suggesting tidal destruction of close-in exoplanets is common. The process can explain the observed cut-off in small *a*-values, the clustering of orbital periods near three days, and the relative youth of transiting planets. Contrary to previous considerations, a mechanism to stop the inward migration of close-in planets at their current orbits is not necessarily required. Planets nearing tidal destruction may be found with extremely small *a*, possibly already stripped of any gaseous envelope. The recently discovered CoRoT-Exo-7 b may be an example of such a planet and will probably be destroyed by tides within the next few Gyrs. Also, where one or more planets have already been accreted, a star may exhibit an unusual composition and/or spin rate.

## 1. Motivation and Background

The distribution of the orbits of close-in exoplanets provides important constraints on models of planet formation and evolution. These planets are unlikely to have formed in their current orbits (with semi-major axis $a < 0.1$ AU) because the protoplanetary gas disk from which they accreted was probably too warm so close to the host star (Lin et al. 1996). For example, the core-accretion model requires coagulation of solid material into a planetary core (Pollack et al. 1996). Within a few 0.01 AU of a star, however, temperatures are too high for condensation of solid material. Instead, the exoplanets observed today in close-in orbits likely formed several AU from their host stars, where temperatures were low enough for condensation of solid material, and later migrated inward (e.g. Lin & Papaloizou 1985). An important process driving the migration to ~ 0.1 AU was probably gas drag in the protoplanetary gas nebula (Lin et al. 1996). Once planets are inside 0.1 AU, tides can play an important role (Rasio & Ford 1996; Weidenschilling & Marzari 1996; Barnes et al. 2008; Jackson et al. 2008a, b; Nagasawa et al. 2008). Here we consider the role of tides in shaping the observed orbital distribution.

Figure 1 shows the semi-major axes and estimated ages for many close-in exoplanets. (The data for Figure 1 are given in Table 1.) Most of these planetary systems are younger than 6 Gyrs, and $a$-values are concentrated between 0.03 and 0.7 AU, centered on about 0.05 AU, which corresponds to an orbital period of about 3 days for a star of one solar mass. Older planets tend to be farther away from their host stars, although there is considerable uncertainty regarding stellar ages (e.g. Saffe et al. 2005; Soderblom 2009) and the sample size is limited. Also, transiting planets (filled squares in Figure 1) tend to be younger than non-transiting planets: all but one transiting planet (XO-5 b) are younger than 6 Gyrs, while there are many older non-transiting planets. Perhaps most significant, there are no planets with $a < 0.017$ AU, and this lower cut-off (approximated by the dashed line in Figure 1) increases with age. For example, among planets younger than 5 Gyrs, *a*-values are as small as 0.017 AU, whereas among planets older than 5 Gyrs, *a*-values are all greater than 0.036 AU. Given the uncertainty in stellar ages, we performed Monte Carlo simulations that incorporated uncertainties in age and *a* and found that this trend is persistent. As we will show, these trends are all consistent with the evolution and eventual destruction of planets by tides.

Previous studies have addressed the apparent concentration of *a*-values centered at 0.05 AU. This concentration is often called the “three-day pile-up” (e.g. Rasio & Ford 2006; Cumming et al. 2008; Jackson et al. 2008a), a term that reflects an implicit assumption that inward migration deposited these planets at their current orbits, like rocks deposited in a moraine at the front edge of a glacier. Thus, studies of the inward migration have tried to account for a stopping mechanism at around 0.05 AU.

Models in which interactions between planets and a protoplanetary gas disk caused the planets to migrate often invoke a clearing of the gas close to the star to halt the migration (Lin et al. 1996; Trilling et al. 1998; Ward 1997; Papaloizou 2007). These studies have suggested a host star’s magnetosphere can clear out the gas disk within a few 0.01 AU of the star, and once the migrating planets enter the cleared region, they stop migrating, resulting in a “pile-up”.

Other studies have invoked a combination of gravitational scattering and tidal evolution to move planets inward to the “pile-up” (Rasio & Ford 1996; Weidenschilling & Marzari 1996; Ford & Rasio 2006; Nagasawa et al. 2008). In such scenarios, interactions between planets in a dynamically unstable system scatter one of the planets into a highly eccentric orbit, with a pericenter distance close enough to the star that tides would affect the orbit. Ford & Rasio (2006) showed that, if scattering resulted in $e \sim 1$ and pericenter distances close to the Roche limit of the star, tides raised on the planet by the host star would drive $a$-values to nearly twice the Roche limit as the eccentricity dropped to zero, at which point tidal evolution resulting only from tides raised on the planet would cease. This process would typically deposit a planet close to 0.05 AU.

However, such models have generally ignored the effect of the tide raised on the host star, which plays a significant role in the evolution of the orbits of close-in exoplanets (Trilling et al. 1998; Dobbs-Dixon et al. 2004; Adams & Laughlin 2006; Jackson et al. 2008a; Levrard et al. 2009; Barker & Ogilvie 2009). In this paper, we consider the effect of the tides raised on the star, which can dominate the orbital evolution for $a < 0.05$ AU. We find that the concentration of observed planets near 0.05 AU may represent the result of inward migration and destruction of closer-in planets, rather than a pile-up of planets that migrated in from farther out. Previous theoretical studies have shown that tides may pull planets into stars in only a few Gyrs (Mardling & Lin 2004; Raymond et al. 2008; Levrard et al. 2009). In this study, we present observational evidence that this destruction has already occurred.

By modeling the tidal evolution of a hypothetical population of close-in planets, we show how the distribution of observed $a$-values with age provides evidence for tidal destruction. We find that many of the features of the observed distribution agree with our models if the ages of our model planetary systems are distributed uniformly. However, in the solar neighborhood, there are observed to be more younger than older stars, and the agreement between the observed and modeled distributions of *a*-values is significantly improved if we account for the prevalence of younger planetary systems. We also find that our model can reproduce the observed clustering of *a*-values if we assume the distribution of initial *a*-values was weighted to smaller *a*.

As we demonstrate below, the tendencies for older exoplanets to be farther from their host stars and for transiting planets to be young likely result from tidal evolution of orbits. We also show that the apparent clustering of orbital periods near three days does not require a mechanism to stop the inward migration of close-in exoplanets. Rather the observed cut-off in *a*-values is a natural result of tidal evolution. Our results indicate that many observed close-in exoplanets will probably be destroyed by tides in a few billion years.

## 2. Tidal Theory

Tides have been considered in many contexts in the study of extra-solar systems. For example, tidal heating may be sufficient to account for many of the discrepancies between the predicted and observed radii of transiting exoplanets (Bodenheimer et al. 2001; Bodenheimer et al. 2003; Burrows et al. 2007; Jackson et al. 2008b; Liu et al. 2008; Ibgui & Burrows 2009; Miller et al. 2009). Tides also reduce orbital eccentricities *e*, probably leading to the observed dichotomy between the relatively narrow distribution of *e*-values for close-in exoplanets and the broad distribution of *e*-values for planets far from their host stars (Rasio et al. 1996; Jackson et al. 2008a).

Tides also change orbital semi-major axes *a* (Jackson et al. 2008a; Levrard et al. 2009), and thus play a key role in producing the observed distribution of *a*-values. As we show below, tides can reduce *a*-values so much that many exoplanets have probably already crossed the critical distance (e.g. the Roche limit) inside of which there would be tidal disruption and accretion of the planet by the host star. As shown below, this tidal destruction probably leads to the observational trends pointed out in Figure 1.

In order to examine how tides can shape the observed distribution (Figure 1), we invoke the classical equations of tidal evolution (c.f. Goldreich 1963; Goldreich & Soter 1966; Jackson et al. 2008; Ferraz-Mello et al. 2008):

$$\frac{1}{a}\frac{da}{dt} = -\left(\frac{63}{2}(GM_*^3)^{1/2}\frac{R_p^5}{Q'_p M_p}e^2 + \frac{9}{2}(G/M_*)^{1/2}\frac{R_*^5 M_p}{Q'_*}\left(1+\frac{57}{4}e^2\right)\right)a^{-13/2} \qquad (1)$$

$$\frac{1}{e}\frac{de}{dt} = -\left(\frac{63}{4}(GM_*^3)^{1/2}\frac{R_p^5}{Q'_p M_p} + \frac{225}{16}(G/M_*)^{1/2}\frac{R_*^5 M_p}{Q'_*}\right)a^{-13/2} \qquad (2)$$

where $G$ is the gravitational constant, $R$ is a body's radius, $M$ its mass and $Q'$ its modified tidal dissipation parameter (as defined in Goldreich & Soter 1966), and subscripts $p$ and $*$ refer to the planet and star, respectively. Note that Equations (1) and (2) include corrections to those presented by Jackson et al. (2008a). They include an additional term of order $e^2$ in Equation (1) due to tides raised on the star and a corrected numerical coefficient in the last term of Equation (2) (e.g. Ferraz-Mello et al. 2008). (Neither of these corrections is great enough to significantly affect the results in Jackson et al. [2008a, b, c, d] or Barnes et al. [2008].)

These equations involve several assumptions about tidal processes, many of which are detailed in Jackson et al. (2008a). One that is particularly relevant here is the assumption that the period of rotation for exoplanetary host stars is always significantly longer than the planet's orbital period, an assumption that is probably reasonable for most close-in exoplanets (see Section 4.2). This assumption yields the particular numerical coefficients in the terms containing $Q'_*$ in Equations (1) and (2). Different assumptions about the periods would change the values and signs of these coefficients (Ferraz-Mello et al. 2008).

Several previous studies overestimate the time for a planetary orbit to decay into the star, the so-called "orbital decay timescale". These studies attempted to simplify the tidal evolution equations by calculating the current values of $a/(da/dt)$ and using it to estimate the timescale, essentially assuming $a$ damps exponentially with time. For example, Rasio et al. (1996) estimated that the orbital decay timescale for 51 Peg b was a trillion years, longer than the main sequence lifetime of its host star. However, the strong dependence of $da/dt$ on $a$ (Equation 1) means the ratio $a/(da/dt)$ is not constant, and therefore $a$ does not damp exponentially. Instead, a numerical solution of the coupled Equations (1) and (2) shows that $da/dt$ rapidly increases in magnitude as $a$ drops. The orbit decays more and more rapidly, as was also pointed out by Jackson et al. (2008a) and Levrard et al. (2009). In fact, 51 Peg b's orbit could decay in only a few billion years, as discussed in more detail below.

Presumably once the planet's pericenter distance $p = a(1-e)$ is close enough to its host star, the tidal gravity of the star will disrupt the planet, and the planet will be destroyed. Figure 2 shows the coupled tidal evolution of pericenter $p$ and eccentricity $e$ from their current values forward in time for two close-in exoplanets for a range of stellar dissipation parameters $Q'_*$. The horizontal line near the bottom of each plot is the Roche limit $a_R$ for that system, given by $a_R = (R_p/0.462)\,(M_*/M_p)^{1/3}$ (Faber et al. 2005). The classical Roche limit involves various assumptions about the planet's density, structure, and strength, so the actual distance of tidal break-up may be different depending on physical parameters. However, the exact value for the Roche limit is not important because once a tidally evolving planet gets within a few 0.01 AU of its host star, its orbit evolves so quickly that the timescale for the destruction of the planet is not sensitive to the exact break-up distance.

In Figure 2a, we show the tidal evolution of CoRoT-Exo-7 b, which is a transiting exoplanet with the smallest radius found to date, about 2 Earth radii (exoplanet.eu), indicating it may be a rocky planet. Its current orbital eccentricity has been reported as zero, so $p = a$. As a result, only the tide raised on the star by the planet contributes to its tidal evolution (Equation 1). As $p$ begins to decrease, the rate of decrease accelerates rapidly. For the commonly adopted value $Q'_* = 10^5$ (Matheiu 1994; Lin et al. 1996),

CoRoT-Exo-7 b's orbit crosses the Roche limit in a few tens of millions of years, but even for $Q'_* = 10^7$, CoRoT-Exo-7 b would be doomed in the next few billion years.

For 51 Peg b (Figure 2b), which was the first exoplanet discovered (Mayor & Queloz 1995), recent estimates give $e = 0.013$ (Butler et al. 2006). Again, as shown in Figure 2b, tidal destruction is possible within the lifetime of the star. However, for this planet, the non-zero *e*-value means that the tide raised on the planet contributes slightly to the orbital evolution. As a result, before 51 Peg b plummets to the Roche limit, its eccentricity decreases, and *p* can rise a bit temporarily. While the pericenter distance does not decrease monotonically, it does eventually accelerate toward the star.

These examples show how dependence of $da/dt$ on $a$ can accelerate the removal of a planet. Consequently, it is the initial *a*-value that determines the time before a planet's orbit crosses the Roche limit, and not so much the initial eccentricity. The acceleration of tidal evolution with decreasing *a* means that a population of planets with small initial *a*-values (orbiting similar stars) would have their *a*-values spread apart more quickly than a group with large initial *a*-values. This process must have a strong effect on the density distribution of those planets as a function of *a* and age, and thus can help shape the observed distribution shown in Figure 1.

## 3. Observable Consequences of Tidal Evolution

### 3.1. Effects of Decreasing *a*

We can illustrate this spreading process and its effect on the orbital distribution by considering a hypothetical population of Jupiter-like planets (with $M_p = 1$ Jupiter mass and $R_p = 1$ Jupiter radius) in initial orbits around sun-like stars (with $M_* = 1$ solar mass, $R_*$ = 1 solar radius and $Q'_*$ = $10^6$) with a range of *a*-values uniformly distributed between 0.01 AU (near the Roche limit) and 0.2 AU (beyond which tides have little effect, e.g. Jackson et al. 2008a). We set initial *e*-values to zero, because (as discussed above) they have little effect on the lifetime before the orbits cross the Roche limit. Here we are considering the evolution of many individual planets, each in its own planetary system, rather than many planets in a single system. For each planet, we modeled the tidal evolution forward in time according to Equation 1. (Equation 2 is not relevant in this case of circular orbits). Figure 3 shows the evolution of the population over 15 Gyrs. The closest planets plunge rapidly in toward their stars, while planets farther out hardly evolve. The planets closest to their stars become less densely distributed in *a*-values.

The decrease in the density of planets for smaller *a*-values is quantified in the histogram in Figure 4, which shows the evolution of number density (# of planets/increment in *a*) as a function of semi-major axis, and how this function changes with age. The results shown in Figure 4 are based on calculations similar to Figure 3, but with many more planets: 10,000 planets spread initially from $a$ = 0.01 to 0.2 AU. For Figure 4, planets are binned by *a*-value in bins 0.005 AU wide, and then the number of planets in that bin is normalized by dividing by the total number of planets in the original population, giving the normalized number density. Initially ($t$ = 0), all bins are filled evenly, reflecting the assumed uniform spread in *a*-values. However, as time passes, the innermost bins are rapidly cleared out, the outermost bins are unaffected, and planets from intermediate bins move inward to smaller *a*-values. As suggested by Figure 2, the rate at which bins clear out depends on the value of $Q'_*$, but for any value of $Q'_*$, the innermost bins clear first.

If the initial *a*-values for real planets were uniformly distributed (and by "initial" we mean when tides began to dominate the evolution), we would expect the distribution of planets to be similar to Figure 4, with relatively few in *a*-bins with low densities and more in bins with high densities. For example, according to Figure 4, among planets that are 1 Gyr old, we might expect to observe about twice as many with *a*-values near 0.03 AU as we find with *a*-values near 0.025 AU.

In order to compare these models of the evolution of number density to the observed distribution, we recast Figure 4 as a plot of semi-major axis vs. stellar age, analogous to Figure 1. From Figure 4, we can extract the locus of points with any fixed number density on a plot of semi-major axis vs. age, yielding the solid curves in Figure 5. For example, consider a normalized number density of 0.015 in Figure 4. We see that after 1 Gyr, the histogram bin corresponding to a density of 0.015 is centered on 0.035 AU. After 2 Gyrs, the bin with that same density is centered on 0.04 AU. In order to map the 0.015 density contour in Figure 5, we draw a curve connecting the point at 1 Gyr and 0.035 AU to the point at 2 Gyrs and 0.04 AU (and also to all the intermediate points). This increase with age in the *a*-value corresponding to a fixed number density is reflected by the positive slope of the contours in Figure 5.

If the population represented by these contours were searched observationally, depending on the total number of planets and the limits of completeness of the search, some bins might not be populated in the observational data. That is, we would not expect to have observed planets in the parts of the distribution below a particular limiting density. Thus the contour line for that density should represent a boundary, below which planets would not appear in the observed population. Note, however, that without knowing the actual number of planets or the completeness of the search to date, we cannot say which specific contour should give that boundary.

However, we can test the plausibility of our model (that the low-*a* cut-off is due to tidal migration into the star) by determining whether any contour of the model distribution follows the cut-off in the observed distribution. Figure 5 includes the observed distribution and the apparent cut-off (from Figure 1) for comparison with the model contours. We see that, for all three values of $Q'_*$, the model contours do predict a paucity of planets with small *a*, and the contours do have slopes toward larger *a* with greater age. But none of the contours matches the relatively steep slope of the observed cut-off (dashed line in Figures 1 and 5).

### 3.2. Effects of the Age Distribution

The model above does not include an important selection effect: there are inherently fewer older stars than there are younger stars in the solar neighborhood (Takeda et al. 2007). Figure 6 shows a histogram of stellar ages for a large population of nearby F, G, and K stars, as reported by Takeda et al. (2007). The number of stars in a given age bin declines roughly linearly with age (although stellar ages do remain very uncertain [e.g. Saffe et al. 2005; Soderblom 2009]). We can also see this trend in Figure 1, especially at large *a*-values (where tides have little effect). For this version of our model, we represent this reduction in the number of older stars with a straight-line fit (as shown in Figure 6).We use that function as the relative probability for a star to be a certain age.

We apply this reduction in the number of older stars to the population shown in Figure 4, reducing the number of planets (uniformly in all *a*-bins) in accord with their increasing age. Figure 7 shows the results of modifying simulated population in this way.

The density contours corresponding to Figure 7 are shown on a plot of semi-major axis vs. age in Figure 8. Compared with Figure 5, the shapes of the contours at low *a* are steeper and, for great enough age, the contours become vertical, reflecting the reduction in all *a*-bins of the number of planets with age.

Comparing the slopes of the model contours to the low-*a* cut-off in the observed population, we find good fits for all values of $Q'_*$. For $Q'_*$ = $10^5$, $10^6$, and $10^7$, respectively, there is a reasonable fit for the density contours with values 0.00015, 0.0015 and 0.0045. In addition, the contours fit the reduction in numbers of old planets for larger *a*-values. Thus, incorporating the reduction in the number of older stars shows that observations are consistent with destruction of planets through tidal decay of their orbits.

### 3.3. Effects of the Initial *a* Distribution

The analysis above assumes that planets begin with a uniform distribution in *a*. We also investigate the effects of choosing an alternative distribution of initial *a*-values. For example, consider an initial population of planets whose number density decreases with increasing *a*, as in the linear distribution at t = 0 in Figure 9. Such a distribution might reflect the effects of gas disk migration, for example. Figure 9 shows the evolution of such a population due to tidal variation of *a* and the loss with age (according to Figure 6). As time passes, the number density for the innermost bins drops rapidly, and the density in the bins farther out remains mostly unchanged, just as in previous calculations. Other assumptions about the initial distribution give qualitatively similar results.

Density contours for semi-major axis vs. age are shown in Figure 10 for this case, analogous to Figures 5 and 8. Here contour lines show that the density drops for increasing *a* where *a* > 0.1 AU. At the lower-*a* cut-off, the contours lines follow the slope of the observed distribution (dashed line). In general, the model contours fit the observed distribution reasonably well. The agreement is especially good for $Q'_* = 10^7$, where the model yields a concentration of *a*-values around 0.05 AU, similar to that observed. These results suggest that we can account for the observed clustering of *a*-values and the low-*a* cut-off by invoking an initial distribution that favors closer planets, followed by tidal orbital decay. The concentration at 0.05 AU may not be a pile-up, but rather it could be the result of carving away part of the initial population. The orbits of many planets have decayed into their parent stars, and more will do so in the future.

## 4. Discussion

### 4.1. Transiting Planets

The process of tidal evolution and destruction of planets discussed above has a profound effect on the observed distribution of *a*-values. It may also explain why transiting planets tend to be younger than non-transiting ones. Transit observations favor discovery of planets closer to their host stars. The slope of the lower cut-off in *a* vs. age space (e.g. the dashed line in Figure 1, which we have shown can be explained by tidal evolution) means that the closest-in planets will tend to be youngest.

Transiting planets are also likely to be destroyed quickly by tides. Levrard et al. (2009) pointed out that the orbits of all known transiting exoplanets (except HAT-P-2 b) are likely to survive less than a few billion years. The preferential destruction of transiting planets relative to planets discovered in other ways should be kept in mind during future statistical studies.

4.2. Stellar Rotation Rates

Our results suggest that many planets have already been accreted into their host star. One potentially observable effect would be an increase in stellar spin rate, as a planet's orbital angular momentum is transferred to the star. The change in rotation rate for a star is given by

$$\Delta\Omega_* = \frac{M_p\sqrt{GM_* a(1-e^2)}}{C_* M_* R_*^2} \qquad (3)$$

where $C_*$ is the stellar moment of inertia coefficient. For example, if a Jupiter-like planet in a circular orbit ($e = 0$) with initial $a = 0.05$ AU plunges into a sun-like star, with a typical 30 day period and $C_* \sim 0.1$ (Massarotti 2008), the rotation period is reduced to only ~ 5 days. Because main-sequence stars would otherwise tend to spin down as they age (Skumanich 1972; S. Barnes 2007), an anomalously high spin rate could be evidence for accretion of a tidally destroyed planet. Some anomalously rapid rotators have been found among red giants, and their rapid rotation rates have been taken as evidence for accretion of planets by the stars (Massorotti 2008). Perhaps candidates could also eventually be identified among main sequence stars.

## 4.3. Stellar Compositions

The accretion of planetary material might also change the host star's composition enough to be measurable. Gonzalez (1997) found that planet-hosting stars seemed to be enhanced in metallicity relative to their stellar neighbors. He proposed that this enhancement was due to pollution of the star by accretion of planetary material, which would be consistent with our model. Fischer & Valenti (2005) found a similar trend but argued that metallicity enhancement of stars is primordial and not due to planetary accretion. In order to resolve this issue, it is important to consider metallicities of stars without planets (as well as those with) because they may have had planets in the past, which affected their metallicity. A particularly useful indicator of relatively recent accretion of planetary material may be $^{6}$Li because Li is quickly destroyed by nucleosynthetic processes (Sandquist et al. 2002). In fact, correlating such spectral signatures with unusually fast rotation rates might provide evidence for past destruction and accretion of a planetary companion.

## 4.4. Stripped Planetary Cores

As a planet reaches the Roche limit, any gaseous atmosphere or envelope could be stripped off by tidal action, leaving only a rocky core (Trilling et al. 1998). Once this tidal stripping begins, it may happen quickly. Faber et al. (2005) showed that mass loss can be rapid for a planet in an eccentric orbit that crosses the Roche limit. But many planets reaching the Roche limit may be on circular orbits, and it would be useful to have models that apply in such cases.

As the planet losses mass, the orbital evolution slows (Equation 1), and the Roche limit for the planet shrinks. Thus the remaining planetary core may survive for a long time (Raymond et al. 2008), so it is possible that a population of such bodies may be discovered as observations become more complete. In fact, the potentially rocky exoplanet CoRoT-Exo-7 b may represent the first discovery in this new class of planets.

4.5. Uncertainties and Assumptions

Although the results presented here have important implications, the details are affected by various uncertainties. The large uncertainties in stellar age and the limited size of the observational sample (so far) are important limitations for our analysis. Stellar ages are difficult to determine (e.g. Saffe et al. 2005; Soderblom 2009), so uncertainties are often large, in many cases as large as the derived age itself. However, our Monte Carlo modeling of the data shows that the slope of the low-$a$ cut-off noted in Figure 1 is robust against these uncertainties. As age estimates improve and more close-in exoplanets are found, our results should be revisited.

Our analysis also involves several assumptions that are reasonable, but subject to revision. For example, the tidal model described by Equations (1) and (2) assumes that the rotational period for the host star is much longer than the revolutionary period for the close-in planet. So far, rotational periods for almost all planet-hosting stars are of order 10 days or longer (Trilling 2000; Barnes 2001; S. Barnes 2007), consistent with our assumption. However, if the rotation period of the host star were to become equal to the orbital period as the planet migrates inward, the tidal evolution would cease, and the planet would not be destroyed. This situation could arise for a few transiting planets if tides spin up the stars sufficiently (Levrard et al. [2009]; see also Counselman [1973], Greenberg [1974], and Dobbs-Dixon et al. [2004]). However, loss of rotational angular momentum (e.g. through shedding of stellar wind or magnetic effects) should help to keep the star's rotation slow, so our assumption will hold in most cases.

We have also assumed that interactions among planets are negligible. In some cases, such interactions could affect the rate of tidal decay of orbits. The removal of a planet from a multiplanetary system through tides may also leave a discernible signature on the dynamical configuration of the remaining planets. These are important topics for on-going research.

The equations we used (1 and 2) assume that the star and planet's tidal dissipation parameters are independent of frequency. Alternative assumptions are possible (Zahn 1977; Hut 1981; Ogilvie & Lin 2007; Penev et al. 2007), and may be important for the cases we consider here, where orbital and rotational frequencies may change rapidly. However, the nature of tidal dissipation within planets and stars is still uncertain, and our results can be revised as understanding of the effects of tides improves.

## 5. Conclusions

Tidal decay of close-in exoplanetary orbits can, and probably has, led to the destruction of many close-in exoplanets. Evidence for this destruction comes from the orbital distribution of observed close-in exoplanets: Older planets tend to be farther from their host stars than younger planets; transiting planets tend to be younger than non-transiting planets; the observed orbits show a cut-off in small $a$-values, which increases to larger $a$-values with increasing age. These trends can be explained by tidal evolution and destruction of close-in exoplanets. Tidal decay rates are quantitatively consistent with the observed distribution of planetary semi-major axes and ages for a wide range of tidal dissipation parameters.

While it is widely assumed that some mechanism must stop the inward migration of close-in exoplanets in order to explain the observed orbital distribution, our results show that it is not necessary. Our model also explains the short life expectancies of some of the closest-in planets (e.g. WASP-12 b, CoRoT-Exo-7 b, and OGLE-TR-56 b). They are simply the next in line to be tidally destroyed after the many that have already spiraled down into their stars. Transiting planets especially, because they must pass close to their host stars to be observed in transit, will quickly be destroyed by tides, which explains why transiting planets tend to be relatively young.

Future observations may provide further evidence for tidal destruction of close-in exoplanets. Stars that have recently consumed a close-in planet may have unusual compositions and may rotate more quickly than expected. Tidal stripping of a gaseous planet's mass once the planet nears the Roche limit may strand the rocky cores of gaseous planets with orbits very close to their host stars, creating a distinctive class of planets that may be seen with improving surveys.

As stellar age estimates become more accurate, more close-in exoplanets are discovered, and our understanding of tidal effects improve, these results will need to be revisited. Nevertheless, our results suggest tidal destruction of close-in exoplanets is common, and many of the exoplanets we see today may soon be gone.

Acknowledgements

The authors acknowledge useful conversations with Josh Eisner, Chris Laws, Michael Meyer, Dylan Morgan, Andrew Skemer, and Christa Van Laerhoven. The paper also benefitted from comments by an anonymous referee. This work was supported by a grant from NASA’s Planetary Geology and Geophysics program and a fellowship to Brian Jackson from the NASA Earth and Space Science Fellowship Program.

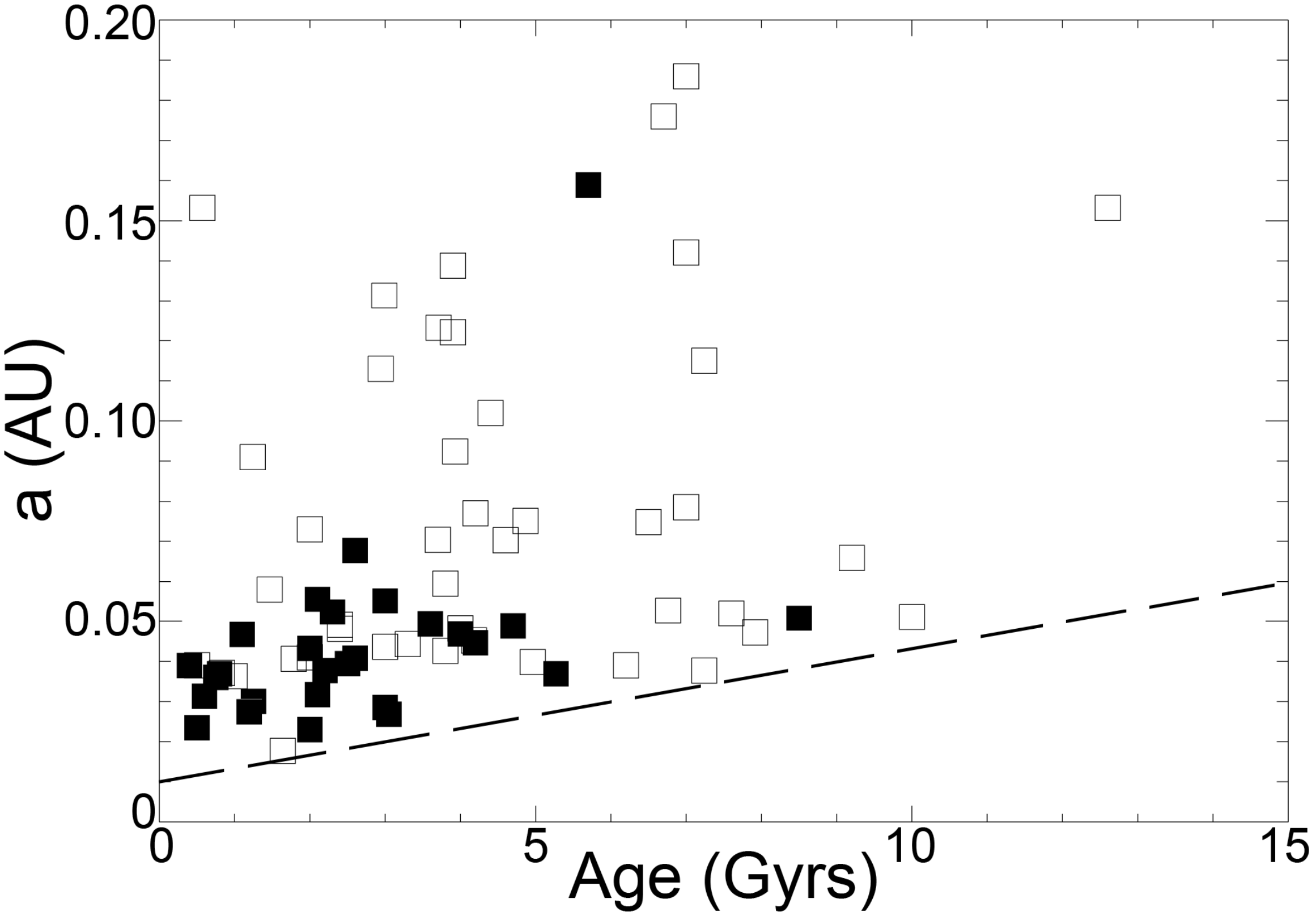

Figure 1: The observed distribution of *a*-values and ages for close-in planets for which some estimate of the age is available. Filled squares represent transiting planets, while open squares represent non-transiting planets. For 9 of the 70 planets in the sample, we only have a lower limit on the age. In those cases, we have plotted the age as the available lower limit. Where only the minimum and maximum values for the age were available, we plotted the average of these values. Sources for the data are listed in Table 1. The dashed line indicates the apparent lower cut-off boundary in observed *a*-values.

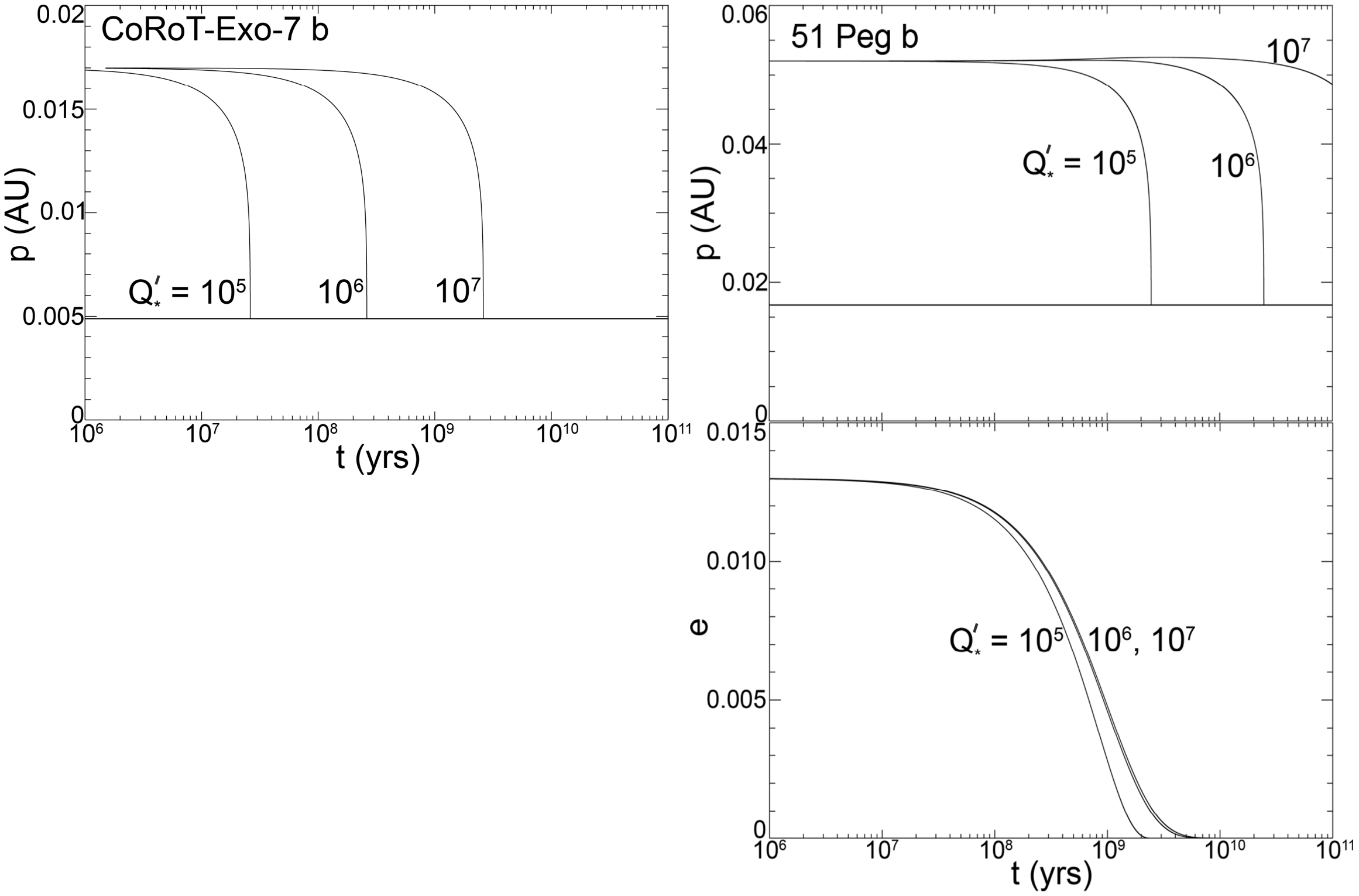

Figure 2: Tidal evolution of the orbits of CoRoT-Exo-7 b and 51 Peg b. (CoRoT-Exo-7 b is reported to have $e$ = 0.) Each line is labeled with its corresponding $Q'_*$ value, and $Q'_p$ is fixed at $10^{6.5}$ (Jackson et al. 2008a). The solid, horizontal line indicates to the Roche limit for each planet and star.

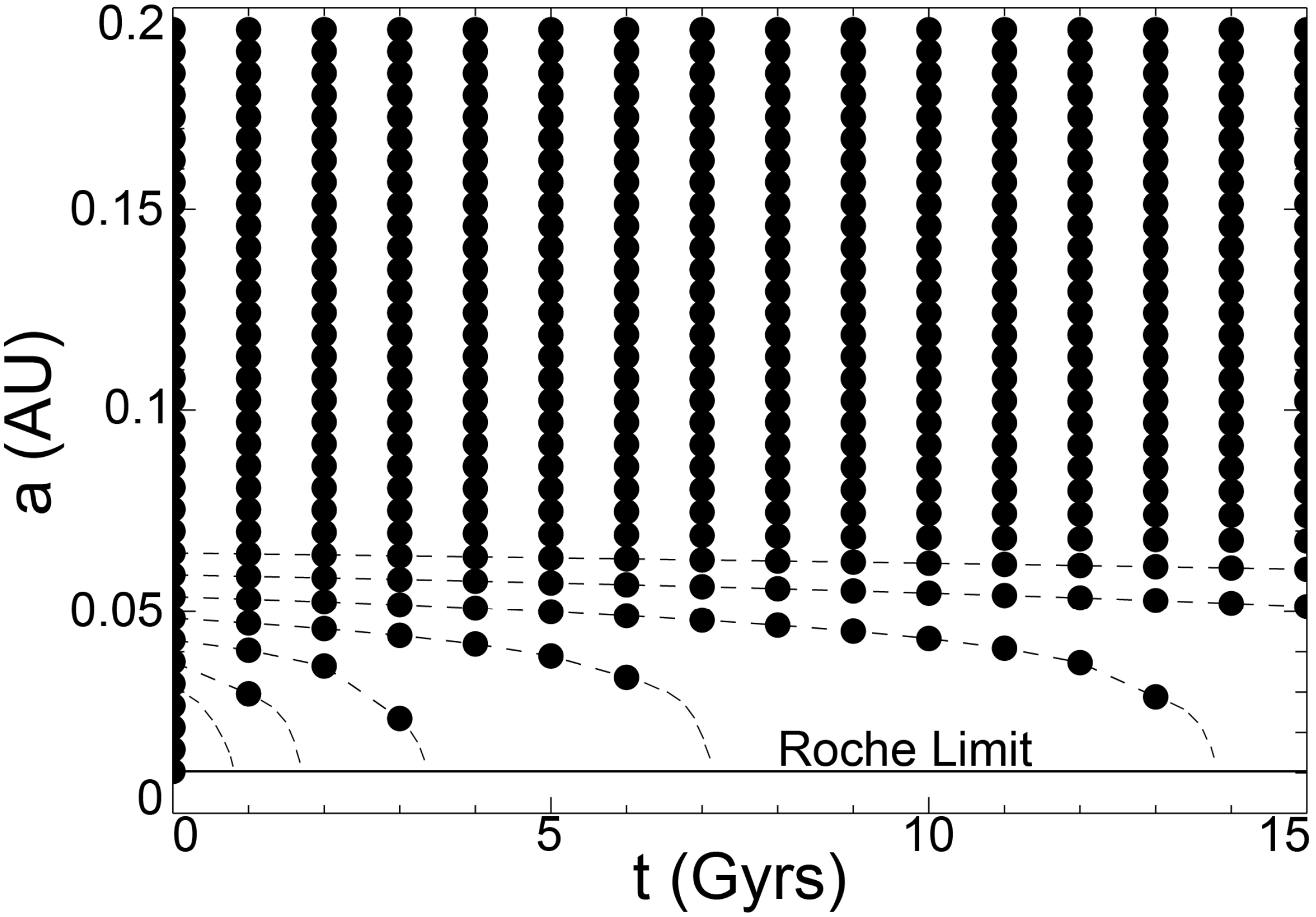

Figure 3: The tidal evolution for a hypothetical population of planets on circular orbits with initial *a*-values uniformly distributed between the Roche limit near 0.01 AU and 0.2 AU. The dashed lines trace the tracks for a few representative examples, and the solid horizontal line shows the Roche limit.

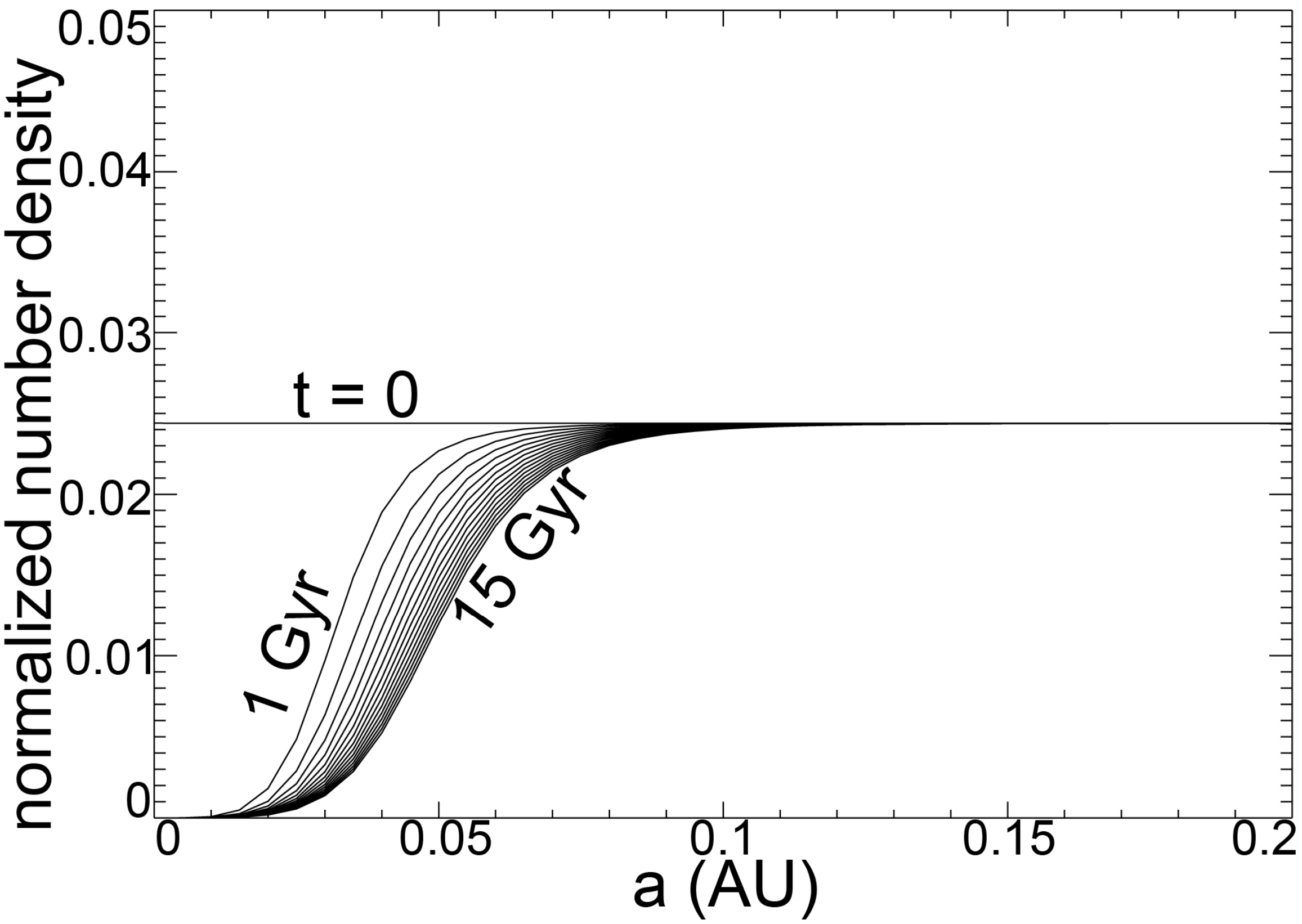

Figure 4: The normalized number density (defined in the text) of a population of planets, shown as a function of *a*, varies over time from an initial uniform distribution. Curves represent the distribution at intervals of 1 Gyr from t = 0 to t = 15 Gyr. For this figure $Q'_*$ = $10^6$.

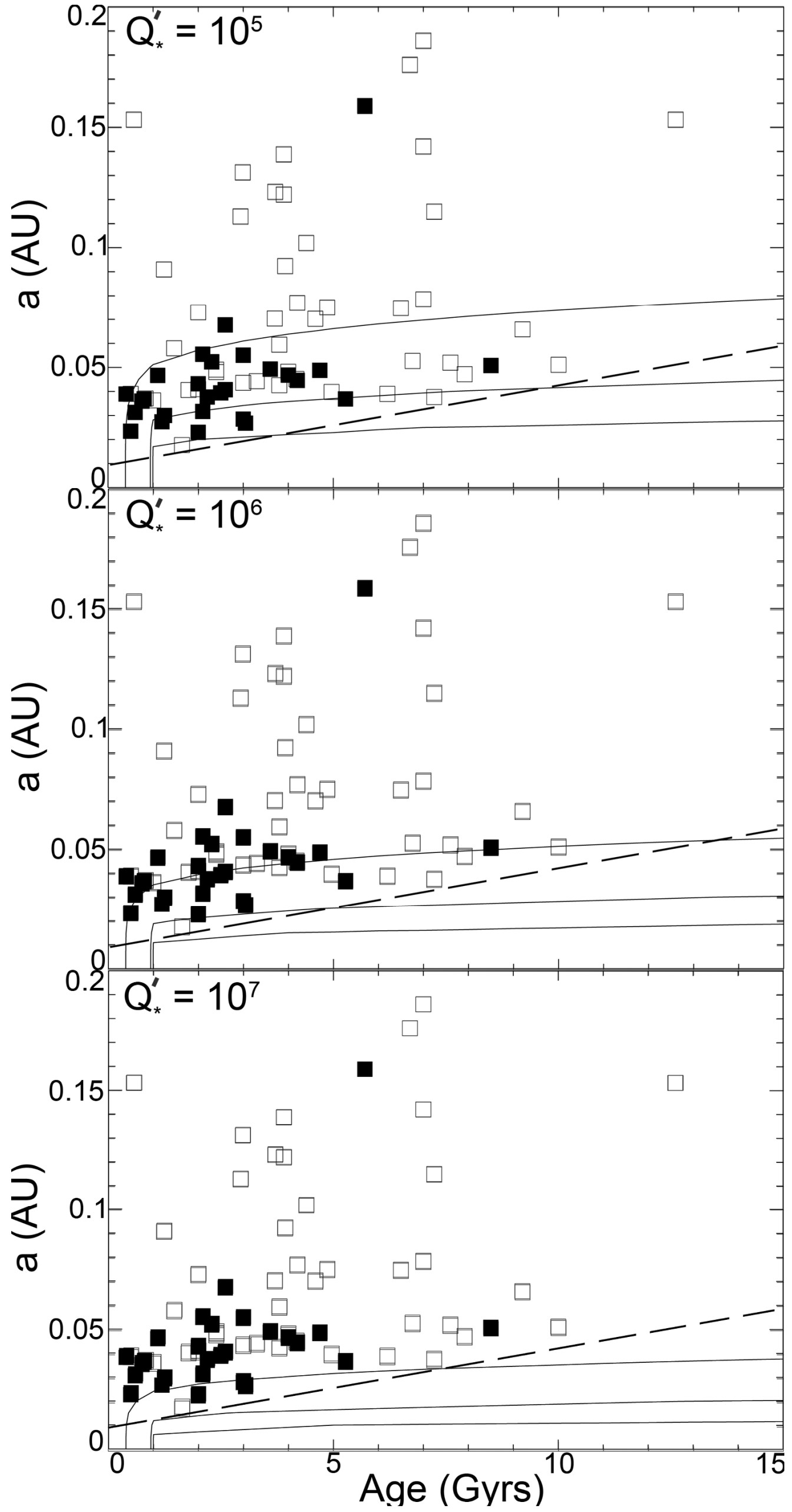

Figure 5: Each solid contour represents the locus of points with a given value for the normalized number density for the population shown in Figure 4. Each panel represents a particular $Q'_*$ value, as labeled. Within each panel, the contours from top to bottom are for 0.00015, 0.0015, and 0.015 in the same units as the y-axis in Figure 4. The squares and dashed line (identical to Figure 1) indicate the observed values and the apparent cut-off in the observed values.

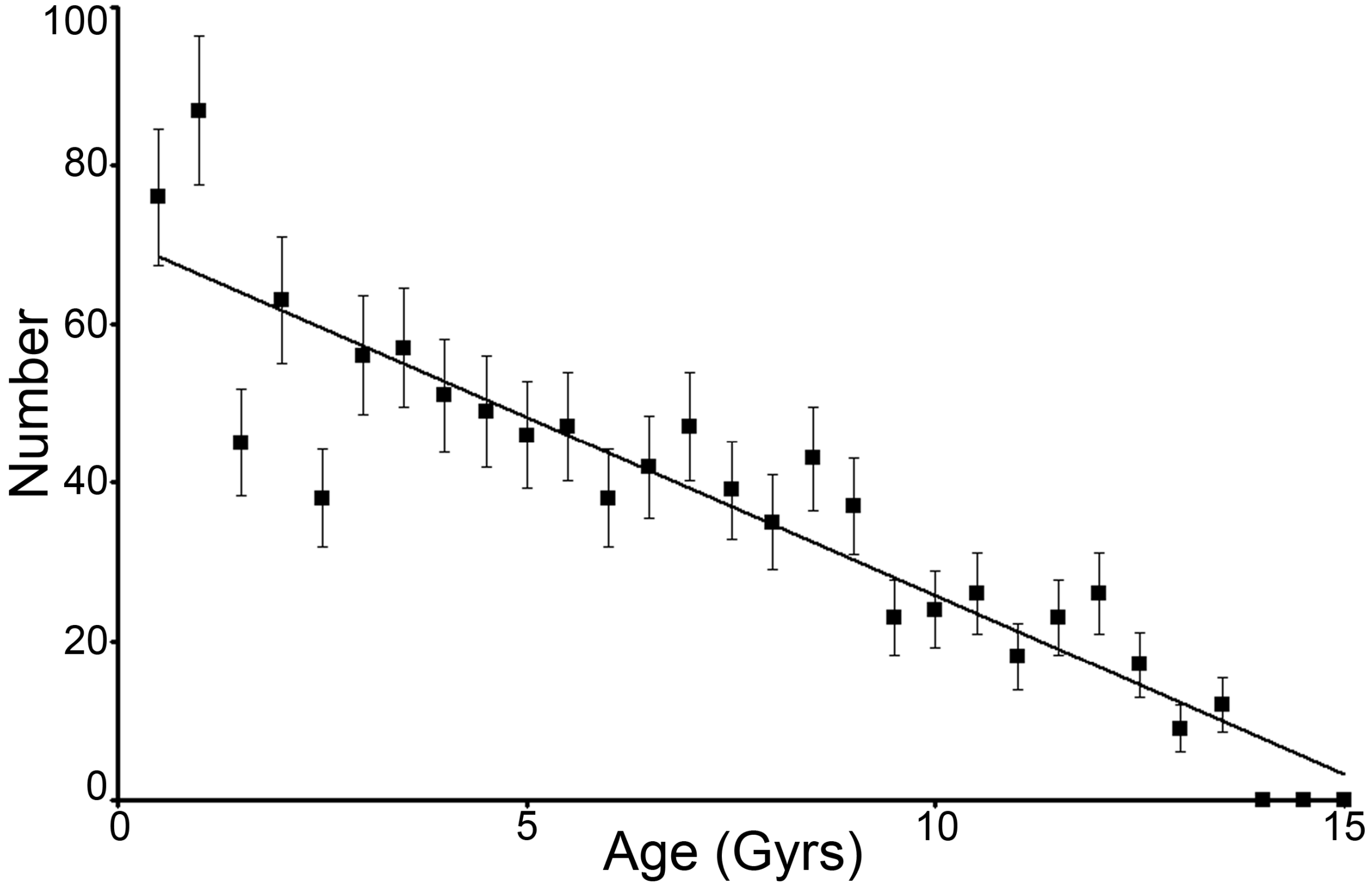

Figure 6: Histogram of stellar ages with a least-squares fit (Takeda et al. 2007). Error bars are based on Poisson statistics.

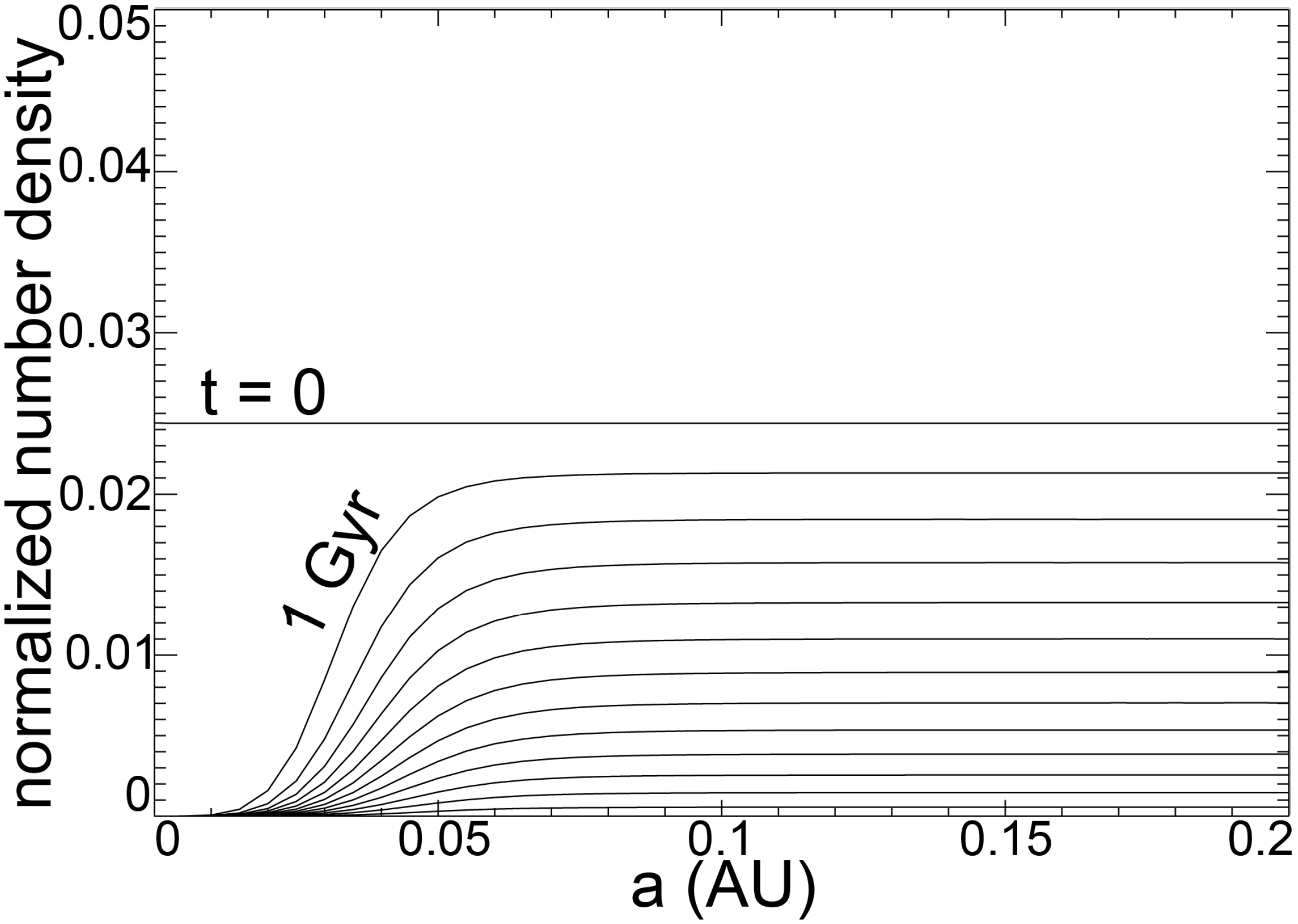

Figure 7: Same as Figure 4, except the number of planets is reduced with time by the same factor in all *a*-bins, according to the linear relationship in Figure 6. For this figure $Q'_* = 10^6$.

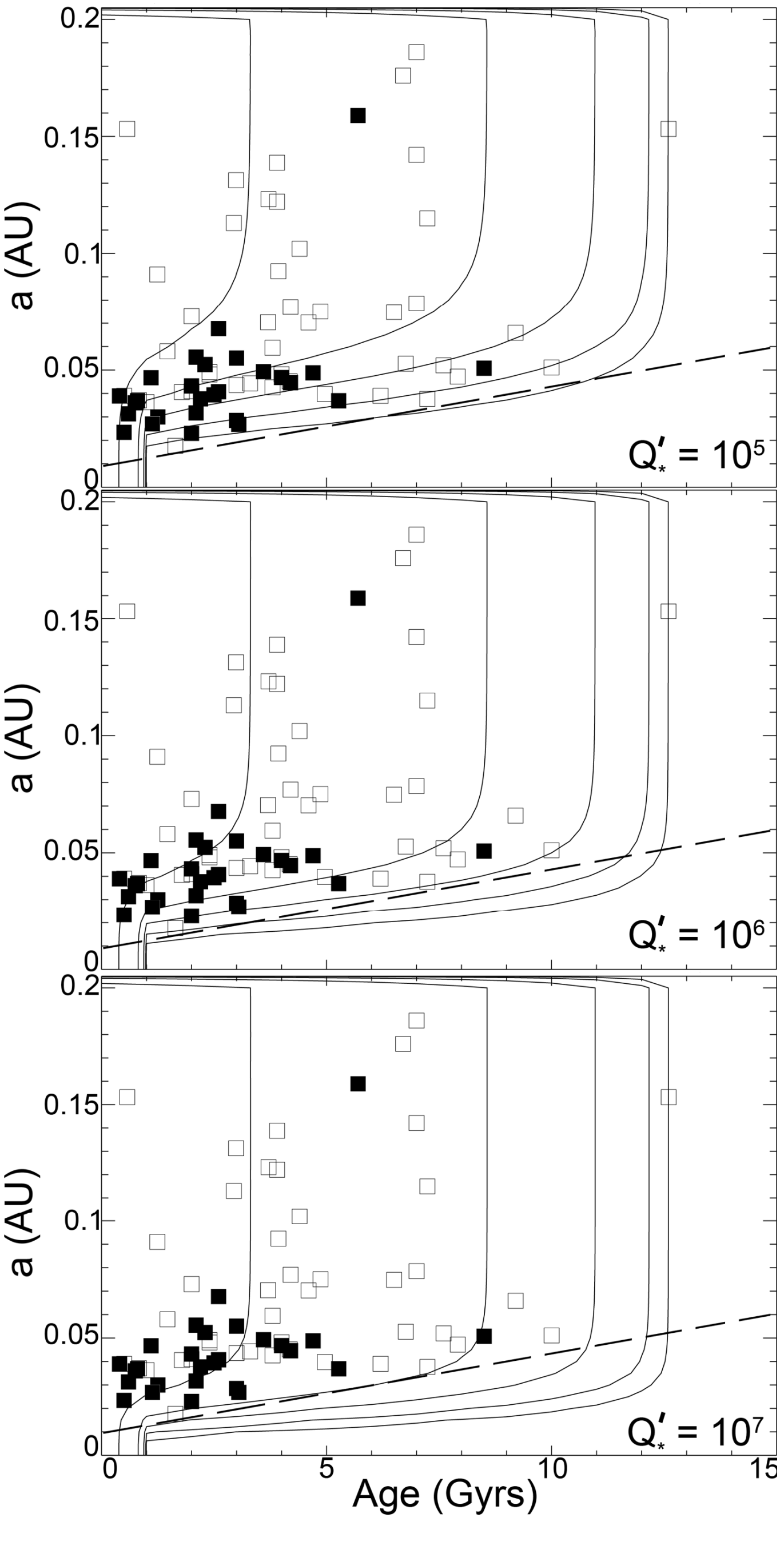

Figure 8: Similar to Figure 5, contours represent the locus of points with a fixed value for the normalized number density for the population shown in Figure 7. Within each panel, contours from bottom right to top left are 0.00015, 0.00045, 0.0015, 0.0045, 0.0015, and 0.045 in the same units as the y-axis of Figure 7. The dashed line and squares have the same meanings as in Figures 1 and 5.

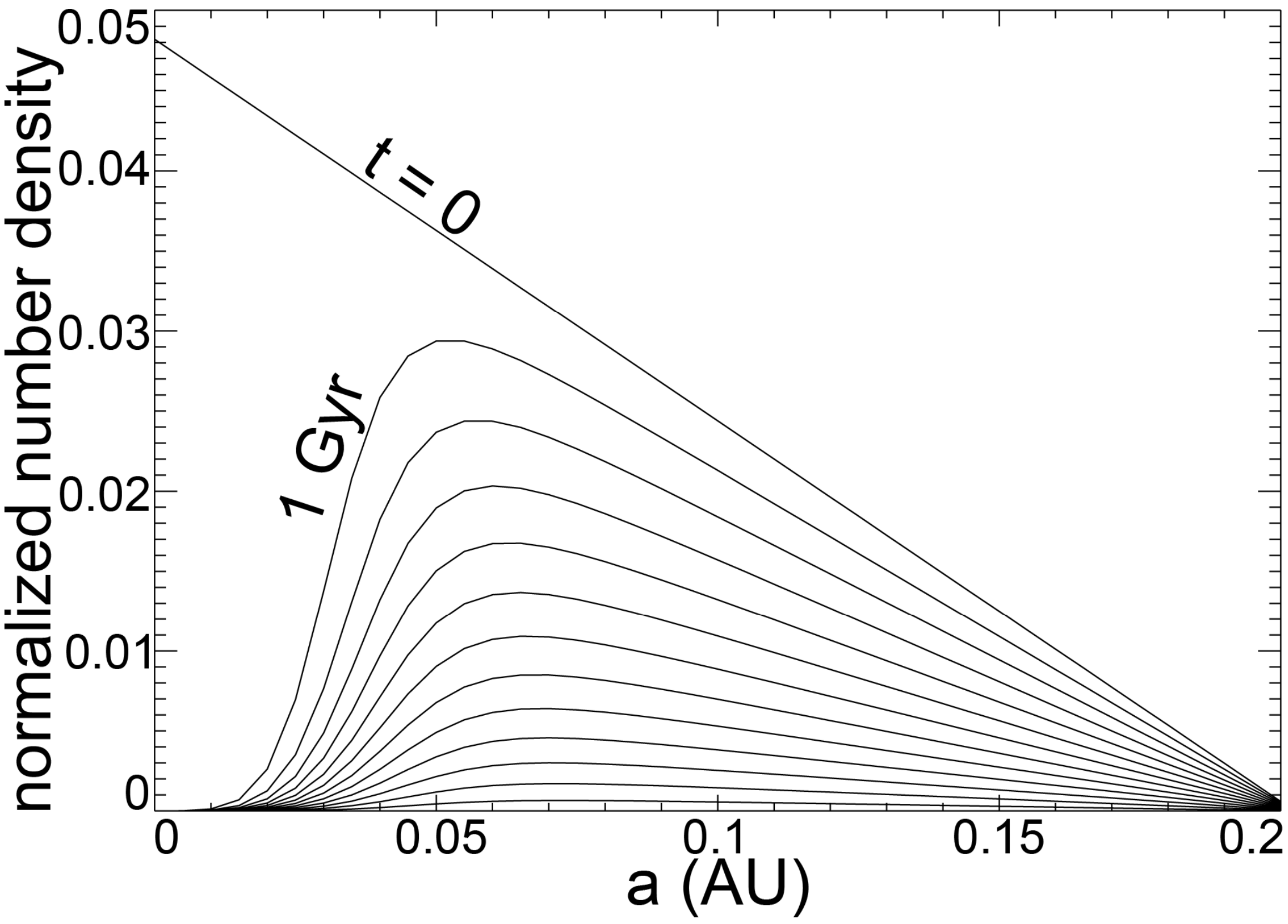

Figure 9: Same as Figure 4, except the number of planets is reduced with time by the same factor in all *a*-bins, according to the linear relationship in Figure 6, and the distribution of initial *a*-values is weighted toward smaller *a*. For this figure $Q'_* = 10^6$.

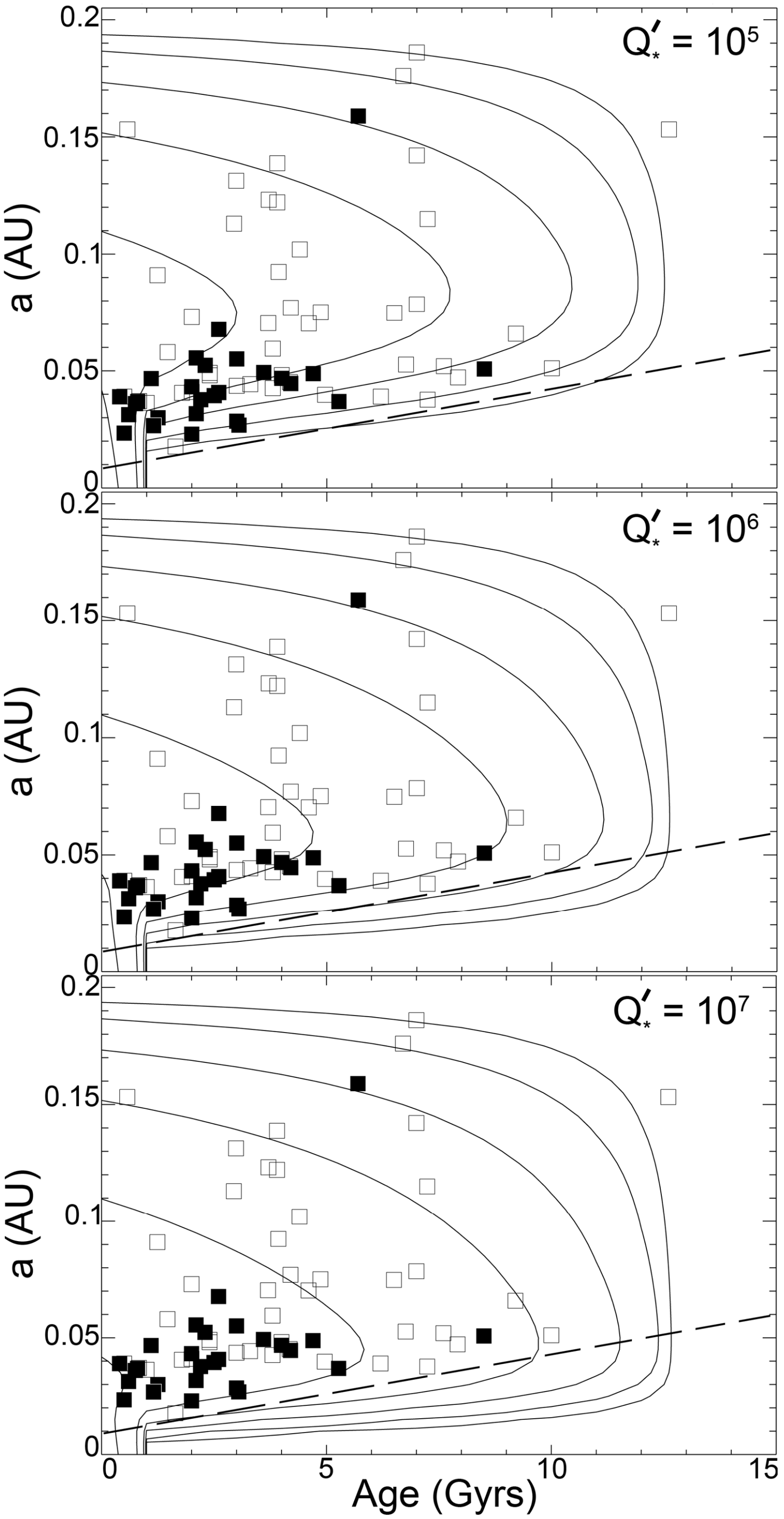

Figure 10: Similar to Figures 5 and 8, contours represent the locus of points with a fixed value for the normalized number density for the population shown in Figure 9. Contours are shown for the same density values as in Figure 8. The dashed lines and squares have the same meanings as in Figures 1, 5, and 8.

| Name | min age (Gyrs) | mid age (Gyrs) | max age (Gyrs) | min *a* (AU) | mid *a* (AU) | max *a* (AU) | age ref, a ref |
|---|---|---|---|---|---|---|---|
| CoRoT-Exo-7 b | 0.7 | 1.1 | 2.2 | … | 0.017 | … | 12, 12 |
| HD 41004 B b | 1.48 | 1.64 | 9.5 | 0.0167 | 0.0177 | 0.0187 | 37, 9 |
| WASP-4 b | 2 | … | … | 0.022 | 0.023 | 0.024 | 42, 42 |
| OGLE-TR-56 b | 0.5 | … | … | 0.023 | 0.0234 | 0.0238 | 36, 36 |
| WASP-5 b | 1.7 | … | 4.4 | 0.0261 | 0.0268 | 0.0277 | 1, 1 |
| GJ 436 b | 3 | … | … | 0.0276 | 0.0285 | 0.0293 | 18, 18 |
| OGLE-TR-132 b | 0.5 | … | 2 | 0.0297 | 0.03 | 0.0302 | 17, 17 |
| HD 189733 b | 0.6 | … | … | … | 0.0313 | … | 29, 43 |
| WASP-3 b | 0.7 | … | 3.5 | 0.0307 | 0.0317 | 0.0322 | 35, 35 |
| WASP-14 b | 0.5 | 0.75 | 1 | 0.035 | 0.036 | 0.037 | 22, 22 |
| HD 63454 b | … | 1 | … | 0.0342 | 0.0363 | 0.0384 | 31, 9 |
| WASP-10 b | 0.6 | 0.8 | 1 | 0.0355 | 0.0369 | 0.0381 | 10, 10 |
| XO-2 b | 2.04 | 5.27 | 9.49 | 0.0367 | 0.0369 | 0.0371 | 7, 7 |
| HD 73256 b | 0.26 | 0.83 | 6.4 | 0.035 | 0.0371 | 0.0392 | 37, 9 |
| 55 Cnc e | 7.24 | … | … | 0.0355 | 0.0377 | 0.0399 | 39, 26 |
| HAT-P-7 b | 1.2 | 2.2 | 3.2 | 0.0372 | 0.0377 | 0.0382 | 33, 33 |
| HAT-P-3 b | 0.1 | 0.4 | 6.9 | 0.0382 | 0.0389 | 0.0396 | 40, 40 |
| GJ 674 b | 0.1 | … | 1 | … | 0.039 | … | 6, 6 |
| HD 330075 b | … | 6.2 | … | … | 0.039 | … | 34, 34 |
| TrES-1 | 1.5 | 2.5 | 3.5 | 0.0394 | 0.0394 | 0.0394 | 29, 43 |
| HD 46375 b | 1.68 | 4.96 | 7.7 | 0.0375 | 0.0398 | 0.0421 | 37, 9 |
| HD 83443 b | 0.63 | … | 2.94 | 0.0383 | 0.0406 | 0.0429 | 37, 9 |
| HAT-P-5 b | 0.8 | 2.6 | 4.4 | 0.04 | 0.0408 | 0.0415 | 4, 4 |
| GJ 581 b | 2 | … | … | … | 0.041 | … | 41, 5 |
| HD 187123 b | 2.26 | 3.8 | 10.6 | 0.0401 | 0.0426 | 0.0451 | 37, 9 |
| HD 149026 b | 1.2 | 2 | 2.8 | 0.0426 | 0.0432 | 0.0438 | 38, 45 |
| HD 2638 b | … | 3 | … | 0.0411 | 0.0436 | 0.0461 | 31, 9 |
| HD 179949 b | 0.4 | 3.3 | 5.4 | 0.0417 | 0.0443 | 0.0469 | 37, 9 |
| HAT-P-4 b | 3.6 | 4.2 | 6.8 | 0.0434 | 0.0446 | 0.0458 | 24, 24 |
| BD-10 3166 b | 0.53 | 4.18 | … | 0.0426 | 0.0452 | 0.0478 | 37, 9 |
| OGLE-TR-111 b | 1.1 | … | … | 0.0402 | 0.0467 | 0.0517 | 29, 30 |
| HD 209458 b | 2 | 4 | 4.5 | 0.0459 | 0.0468 | 0.0477 | 29, 23 |
| HD 88133 b | 6.27 | … | 9.56 | 0.0445 | 0.0472 | 0.0499 | 37, 9 |
| τ Boo b | 0.8 | 2.4 | 3.1 | 0.0453 | 0.0481 | 0.0509 | 37, 9 |
| HD 75289 b | 1.29 | 4 | 5.8 | 0.0454 | 0.0482 | 0.051 | 37, 9 |
| TrES-4 | 2.7 | 4.7 | 6.7 | 0.0466 | 0.0488 | 0.051 | 27, 27 |
| HD 102195 b | 0.6 | … | 4.2 | … | 0.0491 | … | 16, 16 |
| XO-1 b | … | 3.6 | … | … | 0.0493 | … | 20, 20 |
| XO-5 b | 7.8 | 8.5 | 9.3 | 0.0503 | 0.0508 | 0.0511 | 8, 8 |
| HD 76700 b | 0.77 | 10 | 13.1 | 0.0481 | 0.0511 | 0.0541 | 37, 9 |
| HD 149143 b | 6.4 | 7.6 | 8.8 | … | 0.052 | … | 11, 11 |
| HAT-P-6 b | 1.6 | 2.3 | 2.8 | 0.0515 | 0.0524 | 0.0532 | 32, 32 |
| 51 Peg b | 5.28 | 6.76 | 8.4 | 0.0497 | 0.0527 | 0.0557 | 39, 9 |
| HAT-P-1 b | … | 3 | … | 0.0536 | 0.0551 | 0.0566 | 2, 2 |
| XO-4 b | 0.6 | 2.1 | 3.5 | 0.0544 | 0.0555 | 0.0566 | 28, 28 |
| HD 49674 b | 0.55 | … | 2.38 | 0.0546 | 0.058 | 0.0614 | 37, 9 |
| υ And b | 2.8 | 3.8 | 4.8 | 0.0561 | 0.0595 | 0.0629 | 15, 9 |
| HD 168746 b | 3.18 | 9.2 | 10.8 | 0.0621 | 0.0659 | 0.0697 | 37, 9 |
| HAT-P-2 b | 1.2 | 2.6 | 3.4 | 0.0663 | 0.0677 | 0.0691 | 3, 3 |
| HD 118203 b | 3.8 | 4.6 | 5.4 | 0.0662 | 0.0703 | 0.0744 | 11, 9 |
| HD 68988 b | 1.34 | 3.7 | 6.78 | 0.063 | 0.0704 | 0.0745 | 37, 9 |
| GJ 581 c | 2 | … | … | … | 0.073 | … | 41, 5 |
| HD 217107 b | 1.4 | 6.5 | 7.32 | 0.0705 | 0.0748 | 0.0791 | 37, 9 |
| HD 162020 b | 0.23 | … | 9.5 | 0.0708 | 0.0751 | 0.0794 | 37, 9 |
| HD 185269 b | … | 4.2 | … | … | 0.077 | … | 21, 21 |
| HD 69830 b | 4 | … | 10 | … | 0.0785 | … | 25, 25 |
| HD 130322 b | 0.77 | 1.24 | 6 | 0.0857 | 0.091 | 0.0963 | 37, 9 |
| μ Ara d | 1.45 | … | 6.41 | 0.0871 | 0.0924 | 0.0977 | 37, 9 |
| HD 108147 b | 2.3 | 4.4 | 6.6 | 0.0961 | 0.102 | 0.1079 | 37, 9 |
| HD 13445 b | 2.03 | 2.94 | 12.5 | 0.1065 | 0.113 | 0.1195 | 13, 9 |
| 55 Cnc b | 7.24 | … | … | 0.112 | 0.115 | 0.118 | 39, 26 |
| HD 27894 b | … | 3.9 | … | 0.115 | 0.1221 | 0.1292 | 31, 9 |
| HD 99492 b | 2.93 | … | 4.49 | 0.1161 | 0.1232 | 0.1303 | 37, 9 |
| HD 38529 b | 0.89 | … | 5.09 | 0.1237 | 0.1313 | 0.1389 | 37, 9 |
| HD 195019 b | 2.58 | 3.9 | 11.8 | 0.1308 | 0.1388 | 0.1468 | 37, 9 |
| HD 6434 b | 6.85 | 7 | 13.3 | 0.1339 | 0.1421 | 0.1503 | 37, 9 |
| HD 102117 b | 10.9 | 12.6 | 14.3 | 0.1444 | 0.1532 | 0.162 | 37, 9 |
| HD 192263 b | 0.55 | 0.57 | 7.6 | 0.1444 | 0.1532 | 0.162 | 37, 9 |
| HD 17156 b | 3.8 | 5.7 | 7 | 0.1545 | 0.1589 | 0.1643 | 14, 19 |
| HD 117618 b | 3.6 | 6.7 | 9.6 | 0.175 | 0.176 | 0.177 | 37, 9 |
| HD 69830 c | 4 | … | 10 | .. | 0.186 | .. | 25, 25 |

Table 1: Ages and *a*-values of close-in exoplanets used in our analysis. The minimum, middle and maximum values allowed by uncertainties are indicated by “min”, “mid” and “max” respectively. Ellipses indicate values that are unavailable in the literature. The references for ages and *a*-values are given in the rightmost column. References. (1) Anderson et al. (2008), (2) Bakos et al. (2007a), (3) Bakos et al. (2007b), (4) Bakos et al. (2007c), (5) Bonfils et al. (2005), (6) Bonfils et al. (2007), (7) Burke et al. (2007), (8) Burke et al. (2008), (9) Butler et al. (2006), (10) Christian et al. (2008), (11) Da Silva et al. (2006), (12) exoplanet.eu (as of 2009 Feb 10), (13) Fischer & Valenti (2005), (14) Fischer et al. (2007), (15) Fuhrmann et al. (1998), (16) Ge et al. (2006), (17) Gillon et al. (2007a), (18) Gillon et al. (2007b), (19) Gillon et al. (2008), (20) Holman et al. (2006), (21) Johnson et al. (2006), (22) Joshi et al. (2008), (23) Knutson et al. (2007), (24) Kovacs et al. (2007), (25) Lovis et al. (2006), (26) MacArthur et al. (2004), (27) Mandushev et al. (2007), (28) McCullough et al. (2008), (29) Melo et al. (2006), (30) Minniti et al. (2007), (31) Moutou et al. (2006), (32) Noyes et al. (2008), (33) Pal et al. (2008), (34) Pepe et al. (2004), (35) Pollaco et al. (2008), (36) Pont et al. (2007), (37) Saffe et al. (2006), (38) Sato et al. (2005), (39) Takeda et al. (2007), (40) Torres et al. (2007), (41) Udry et al. (2007), (42) Wilson et al. (2008), (43) Winn et al. (2007a), (44) Winn et al. (2007b), (45) Winn et al. (2008).